\documentclass{aa}
\usepackage{graphicx}
\begin{document}
\title{Chemical composition of Galactic OB stars}
\subtitle{II. The fast rotator \object{$\zeta$ Oph}\thanks{The INT is operated on the island of La Palma by the ING in the Spanish Obervatorio de El Roque de los Muchachos of the Instituto de Astrof\'\i sica de Canarias.}}

\author{M.R. Villamariz\inst{1}
          \and
          A. Herrero\inst{2}
          }

\offprints{M.R. Villamariz}

\institute{Instituto de Astrof\'\i sica de Canarias, E-38200 La Laguna, 
	   Tenerife, Spain\\
             \email{ccid@ll.iac.es}
           \and
           Instituto de Astrof\'\i sica de Canarias, E-38200 La Laguna, 
	   Tenerife, Spain\\
	   Departamento de Astrof\'{\i}sica, Universidad de La Laguna,
	   E-38071 La Laguna, Tenerife, Spain\\
              \email{ahd@ll.iac.es}
             }

\date{Received 9 February 2005; Accepted 20 May 2005}


\abstract{\object{$\zeta$ Oph}, \object{HD\,149757}, is an O9.5 Vnn star with a very high projected rotational velocity (v{\thinspace}sin{\thinspace}$i \geq $340 km\,s$^{-1}$). It is also a classical runaway star due to its high proper motion. We perform a quantitative analysis of its optical spectrum in order to measure important observables of the star such as its mass, effective temperature, luminosity and He, C, N, and O abundances. Comparing these observed values to those predicted by the rotating evolutionary models of the Geneva group we find that none of the two sets of models is capable of reproducing the characteristics of the star. Nevertheless, due to its runaway nature, the reason for this discrepancy may be that the star is not the result of the evolution of a single object, but the product of the evolution of a close binary system. 
		}
\maketitle


\section{Introduction}

\object{HD\,149757}, \object{$\zeta$ Oph}, is an O9.5 Vnn star (Mason et al.\cite{mason98}), with a very high projected rotational velocity (v{\thinspace}sin{\thinspace}$i \geq $340 km\,s$^{-1}$ in all published values, as in Penny, \cite{penny96}, Herrero et al., \cite{h92} and Howarth \& Smith \cite{HowyS01}) and a considerable He enhancement in its atmosphere, according to Herrero et al. (\cite{h92}) who measured an He abundance by number of $\epsilon$(He)=0.16\footnote{With $\epsilon$(He)=$\frac{N(He)}{N(H)+N(He)}$; for reference the solar value is 0.08 (Ballantyne et al. \cite{ball00})} for this object.

Recent models of massive stellar structure and evolution with rotation (Meynet \& Maeder, \cite{meymae00}, \cite{meymae03}, Heger \& Langer, \cite{HegyLang00} and references therein) show how stellar rotation can induce mixing processes in the star that bring nuclear processed material from the stellar core to the atmosphere, even at the first stages of the life of the star, during the main sequence phase (MS). This mixing is found to be more efficient for more massive objects and higher rotational velocities, and it is also a decreasing function of the metallicity. The sequence of surface contamination in the MS is the following: first a N enhancement appears, that grows as He enrichment starts, and finally CO depletions also occur (Meynet \& Maeder, \cite{meymae00}, Heger \& Langer, \cite{HegyLang00}).

Our work (Villamariz et al. \cite{cid02}, hereafter paper I) studies quantitatively the He \& CNO abundances for a modest sample of O9 stars, showing that surface CNO contamination certainly exists for the fast rotator of the sample, HD\,191423 (v{\thinspace}sin{\thinspace}$i$=450 km\,s$^{-1}$). Many other authors have also found this effect in some OB stars, mainly in B supergiants (Gies \& Lambert, \cite{GyLambert92}, Lennon et al., \cite{Lennon93}, McErlean et al., \cite{mac99}, Daflon et al. \cite{Daf01}), and Venn (\cite{Venn95}) also for some A supergiants. But a one-to-one relation between stellar rotation and these surface CNO patterns is still not clear.

For an object like \object{HD\,149757}, with such a high rotational velocity, the He enhancement found in Herrero et al. (\cite{h92}) agrees also with the above considerations, but a study of the CNO abundances is mandatory to give a more complete picture. Also the He abundance has to be revised to account for the lower He abundances that we determine now due to improvements in our analysis (see Villamariz \& Herrero \cite{coimbra02} and paper I for details).

Therefore, as the natural progress of our series on the chemical composition of Galactic OB stars we present our study about the He \& CNO abundances of \object{HD\,149757}, that together with other parameters such as its age and mass, will allow us to make a quantitative comparison with the predictions of the evolutionary models.

Nevertheless, the star is a classical runaway (Blaauw \cite{Blaa61}, \cite{Blaa93}, Gies \cite{Gies87}) due to its high proper motion ($\mu_{\alpha}$=13.07 $\pm$ 0.85 mas/yr, $\mu_{\delta}$=25.44 $\pm$ 0.72 mas/yr, from the Hipparcos Catalogue, ESA \cite{ESA1997}). Therefore, it may not be a single star evolving by itself, but the result of the evolution of a close binary system, as considered in the literature (from Blaauw \cite{blaa52} to Hoogerwerf et al., \cite{hoo01}). If this was the case, our discussion in terms of single massive stellar evolution would not be suitable, but the runaway origin of $\zeta$ Oph is not clear in part due to the results of this work and also because of another important question.

The two scenarios accepted nowadays for the formation of massive runaway stars, the BSS (Binary-supernova scenario) and the DES (Dynamical ejection scenario) as explained for example in Hoogerwerf et al. (\cite{hoo01}) need to be reformulated to account for the recent findings of massive stellar evolution with rotation. To explain the observable differences between these two types of objects it is implicitly considered that surface contamination of the runaway can only be due to the accretion of processed material from its primary companion during the evolution of the close binary system, which is not necessarily the case as explained above: self contamination can occur in the isolated evolution of a massive rotating star.

We will address this question in more detail and attempt a definitive formation scenario for $\zeta$ Oph in a forthcoming paper; here we restrict our study to the question of whether the observable characteristics of $\zeta$ Oph are consistent with the evolution of a rotating single massive star.

In sect. \ref{hhe} we present our new analysis of the H/He spectrum, that provides the new He abundance together with other stellar parameters such as effective temperature, logarithmic surface gravity, stellar mass and luminosity. In sect. \ref{cno} we determine the CNO abundances and present the improvements we have introduced in the analysis. In sect. \ref{evol} we compare the abundances to what massive stellar evolution with rotation predicts for this object and finally we conclude in section \ref{discuss}.


\begin{figure}
\centering
\includegraphics[width=9cm]{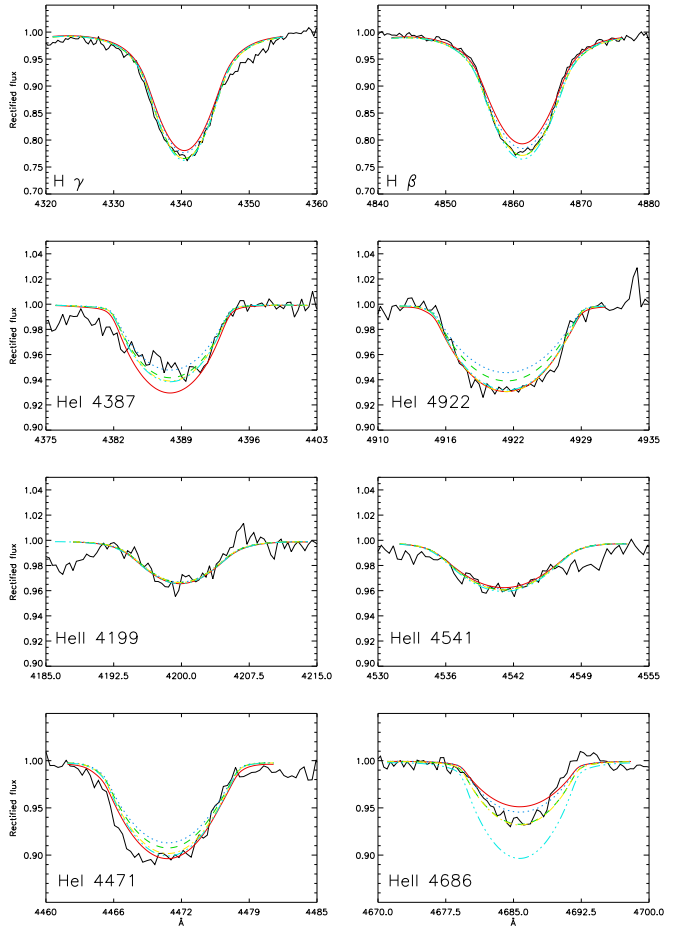}
\caption{Evolution of the parameters of \object{HD\,149757}, from those of Herrero et al. (\cite{h92}, model M1, solid line) to ours (model M5, dash-three dotted). Model M2 (dotted) accounts for the new criterion for \ion{He}{i} lines, M3 (dashed) for {\it line-blocking}, M4 (dash-dotted) for microturbulence and the final M5 (dash-three dotted) accounts for all the previous with the updated version of  {\it line-blocking}. See text for details.
Our uncertainties are the following: $\Delta T_{\mathrm eff}$=1\,000 K, $\Delta \log g$=0.10 dex, $\Delta \epsilon$(He)= 25\% and $\Delta \xi$=2 km\,s$^{-1}$}
\label{149pars}
\end{figure}

\section{New analysis of the H/He spectrum}\label{hhe}

The improvements that we have introduced in our NLTE planeparallel and hydrostatic analyses of OB stellar atmospheres, in comparison to those of Herrero et al. (\cite{h92}), are three. First we adopt a slightly different criterion for the \ion{He}{i} lines suitable for the analysis in the late types (O9 and early B, see Villamariz \cite{tesis} for details). Second is the inclusion of {\it line-blocking} in the spectral synthesis, first as in Herrero et al. (\cite{h00}, lb I in table 1) and later on with our actual better approximation presented in paper I (sect. 3.1, lb II in table 1). The third is the consideration of microturbulence ($\xi$) in the synthesis of the H/He lines (Villamariz \& Herrero, \cite{micro00}).

The observations used for this work are those presented in Herrero et al. (\cite{h92}), with the ING 2.5m Isaac Newton Telescope and the Intermediate Dispersion Spectrograph, with the H2400B grating yielding a spectral resolution of 0.6 \AA~and with exposure times giving a S/N in excess of 200 in the H$_{\rm\beta}$ region.

In Fig. \ref{149pars} we can see the evolution of the stellar parameters of $\zeta$ Oph from Herrero et al. (\cite{h92}, model M1, solid line) to the present ones (model M5, dash-three dotted). The other three model spectra are the best fits considering all the improvements one by one, i.e, first considering only the new criterion for the \ion{He}{i} lines (M2, dotted line), then adding {\it line-blocking} as in Herrero et al. (\cite{h00}, M3, dashed), then considering also microturbulence in the synthesis of the H/He lines (M4, dash-dotted) and finally with the updated {\it line-blocking} of paper I (M5, dash-three dotted).

\begin{table*}[!t]
\label{HHepars}
\caption[ ]{Parameters of the models plotted in Fig. \ref{149pars}, M1 to M5, and of all the models used. Note that for CNO $\epsilon$(X) is in the standard $\epsilon$(X)=12+$\log \frac{X}{H}$, while for He $\epsilon$(He)=$\frac{N(He)}{N(H)+N(He)}$. Our v{\thinspace}sin{\thinspace}$i$=400 $\pm$ 20 km\,s$^{-1}$.}
\begin{center}
\begin{tabular}{cccccccclc} 
\hline
Mod. & $T_{\mathrm eff}$ & $\log g$ & $\epsilon$(He) & $\xi$  & $\epsilon$(C) & $\epsilon$(N) & $\epsilon$(O) & Effect & Line \\
\hline
 M1& 32\,500& 3.55 & 0.16 & 0  & 0 & 0 & 0 & initial & solid \\ 
 M2& 34\,000& 3.70 & 0.09 & 0  & 0 & 0 & 0 & \ion{He}{i} lines & \verb|....| \\ 
 M3& 34\,000& 3.70 & 0.09 & 0  & 0 & 0 & 0 & lb I & \verb|----| \\ 
 M4& 34\,000& 3.70 & 0.09 & 15 & 0 & 0 & 0 & $\xi$ & \verb|.-.-| \\ 
 M5& 34\,000& 3.70 & 0.11 & 15 & 0 & 0 & 0 & lb II & \verb|.-..| \\ 
 M6& 34\,000& 3.70 & 0.11 & 15 & 8.19 & 7.89 & 8.69& --& -- \\
 M7& 34\,000& 3.70 & 0.11 & 15 & 7.86 & 8.34 & 8.69& --& -- \\
 M8& 34\,000& 3.70 & 0.09 & 15 & 7.86 & 8.34 & 8.69& --& -- \\
 M9& 34\,000& 3.70 & 0.09 & 15 & 7.86 & 8.34 & 8.69& --& -- \\
\hline
\end{tabular}
\end{center}
\end{table*}

The value of 15 $\pm$ 2 km\,s$^{-1}$ for microturbulence is the mean value found for the two O9 dwarfs analysed in paper I, for which microturbulence could be determined independently from their \ion{N}{iii} and \ion{C}{iii} lines.

\begin{figure}
\centering
\includegraphics[width=9cm]{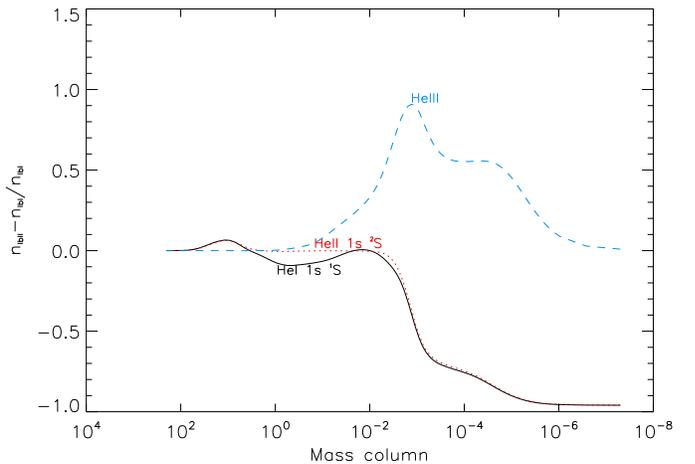}
\caption{Comparison of the populations of the ground levels of \ion{He}{i} (solid), \ion{He}{ii} (dotted) and \ion{He}{iii} (dashed) for the two treatments of {\it line-blocking} (lbI and lbII, see text). The formation depth of all the relevant He lines is above mass column 1.}
\label{pops}
\end{figure}

In all the references quoted at the beginning of the section it is shown how the new He abundances determined are systematically lower than or equal to the previous ones, the greater differences being for the hottest and more extended objects. The only exception to this is the consideration of the more complete {\it line-blocking} of paper I, that as can be seen in Table 1 produces a slightly higher He abundance. This result is in contradiction to what we expected in paper I, where we thought that the new opacities would not affect the formation of the H/He lines, since the strongest contribution to the H/He occupation numbers comes from a spectral range that is common to both treatments of {\it line-blocking}. Nevertheles see in Fig. \ref{pops} how the new treatment produces a (lbII) model with more \ion{He}{i} and \ion{He}{ii} ionized into \ion{He}{iii}. This produces weaker \ion{He}{i} and \ion{He}{ii} lines, that explains the necessity to increase the He abundance in order to fit the observed lines.

Regarding the individual and relative quality of the line fits, the subsequent models reproduce what we find in all previous works. The extremely poor fit to the \ion{He}{ii}\,4686 line with our final model is perfectly acceptable for this line in this spectral range (see discussion on this line in Herrero et al., \cite{h00}, sect. 6.2).

Thus we find that the He abundance of our object is nearly normal or slightly enhanced. As clearly shown in Table 1 this new abundance relies on the new criterion for the \ion{He}{i} lines, which is somewhat subjective. Nevertheless, this is the only possibility, since the simultaneous fitting of all \ion{He}{i} lines with a unique set of stellar parameters is not possible in this spectral range, as many other authors in the literature have also found (see McErlean et al., \cite{mac99}, and references therein).

Howarth \& Smith (\cite{HowyS01}) consider the von Zeipel's variation of $T_{\mathrm eff}$ and $\log g$ with latitude in a rotating object, something that is not usually done. Apart from this remarkable improvement, their model physics and fitting criteria are essentially those of Herrero et al. (\cite{h92}), and therefore their comparable He abundances of 0.20 $\pm$ 0.03 and 0.16 $\pm$ 0.04 respectively confirms that our neglection of the von Zeipel's law in the analysis is not important, even for this very fast rotator.

\subsection{Stellar mass, age and position in the HR diagram}

\begin{figure}
\centering
\includegraphics[width=9cm]{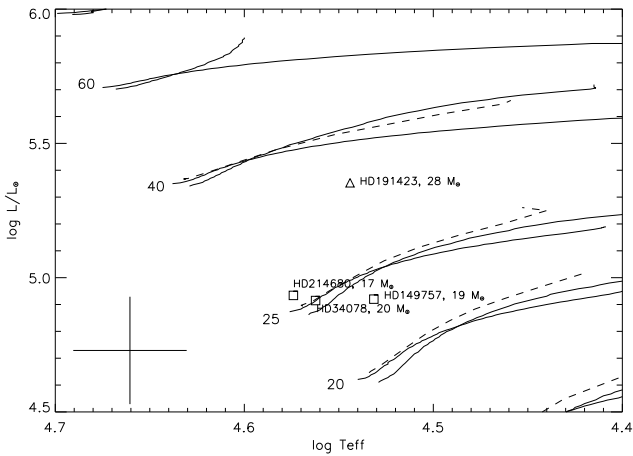}
\caption{HR diagram with evolutionary tracks of Meynet \& Maeder (\cite{meymae00}: dashed and \cite{meymae03}: solid). The new evolutionary tracks are for initial equatorial rotational velocities of 0 and 300 km\,s$^{-1}$ while the older tracks are only for 300 km\,s$^{-1}$. The position of HD\,149757 is shown together with objects we studied in paper I (see Table \ref{O9refs} for details), also their masses are plotted. The cross at the right shows the uncertainties in $T_{\mathrm eff}$ and $log$ L.}
\label{149HRD}
\end{figure}

Directly from the previous analysis, making use of the obtained parameters and adding to them only the stellar absolute visual magnitud (-4.2, Howarth \& Prinja \cite{hp89}), we can obtain a stellar radius (as in Kudritzki et al. \cite{kud80}) of 8.3 $\pm$ 1.5 R$_{\odot}$, a mass of 19 $\pm$ 11 M$_{\odot}$ and a stellar luminosity of 8.3 $\pm$ 4.0 $\cdot10^4$ L$_{\odot}$; we can also place the object in the HR diagram to attempt an age determination (see Fig. \ref{149HRD}).

Interestingly, with their more sophisticated methodology, Howarth \& Smith (\cite{HowyS01}) find a mass of 20 M$_{\odot}$, an area-equivalent radius of 8.5 R$_{\odot}$ and a luminosity of 9.12$\cdot10^4$ L$_{\odot}$, in agreement with our results.

Considering its position in the HR diagram (HRD), $\zeta$ Oph seems to be slightly more evolved than our reference slowly rotating O9 V stars in paper I, HD\,214680 and HD\,34078, who lie very close to the zero age MS (ZAMS, see Fig. \ref{149HRD}). They are reference objects in the sense that they do not show any CNO contamination on their surface.

To estimate the age of our object we intersect the vertical line defined by its effective temperature with the two sets of rotating evolutionary tracks (see section \ref{evol} for details). For the 2000 tracks, there is no intersection with the models for 20 M$_{\odot}$ and initial rotational velocities of 400, 500 and 611 km\,s$^{-1}$. These tracks are not shown in Fig. \ref{149HRD} for clarity, but they are the most appropiate considering the actual value of v{\thinspace}sin{\thinspace}$i$ $\ge$ 340 km\,s$^{-1}$ of our object.

Linearly extrapolating the tracks at 200 and 300  km\,s$^{-1}$ for 25 M$_{\odot}$ up to 500 km\,s$^{-1}$, the age of our object has an upper limit of 5 Myr ($\tau \le$ 5Myr). Doing the same for the 2003 tracks the upper limit shifts to 3.8 Myr.

\begin{table}
\caption[]{Parameters of $\zeta$ Oph and the reference objects analysed in paper I}
\label{O9refs}
  $$ 
\begin{array}{p{0.2\linewidth}cccccc}
\hline
\noalign{\smallskip}
Star & {\mathrm{Clasif.}} & v{\thinspace}\sin{i} & T_{\mathrm{eff}} & \log{g} & \epsilon({\mathrm{He}}) & \xi \\
\noalign{\smallskip}
\hline
\object{HD\,149757} & {\mathrm{O9.5\,Vnn}} & 400 & 34\,000 & 3.70 & 0.11 & 15 \\
\noalign{\smallskip}
\object{HD\,191423} & {\mathrm{O9\,III:n}} & 450 & 35\,000 & 3.40 & 0.12 & 20 \\
\object{HD\,214680} & {\mathrm{O9\,V}}     & 50  & 37\,500 & 4.00 & 0.10 & 15 \\ 
\object{HD\,34078}  & {\mathrm{O9.5\,V}}   & 40  & 36\,500 & 4.05 & 0.09 & 15 \\ 
\noalign{\smallskip}
\hline
\end{array}
   $$ 
\end{table}


\section{Analysis of the CNO spectrum}\label{cno}

\subsection{Preliminar comparisons}
\begin{figure*}[!ht]
\centering
\includegraphics[width=8cm,angle=90]{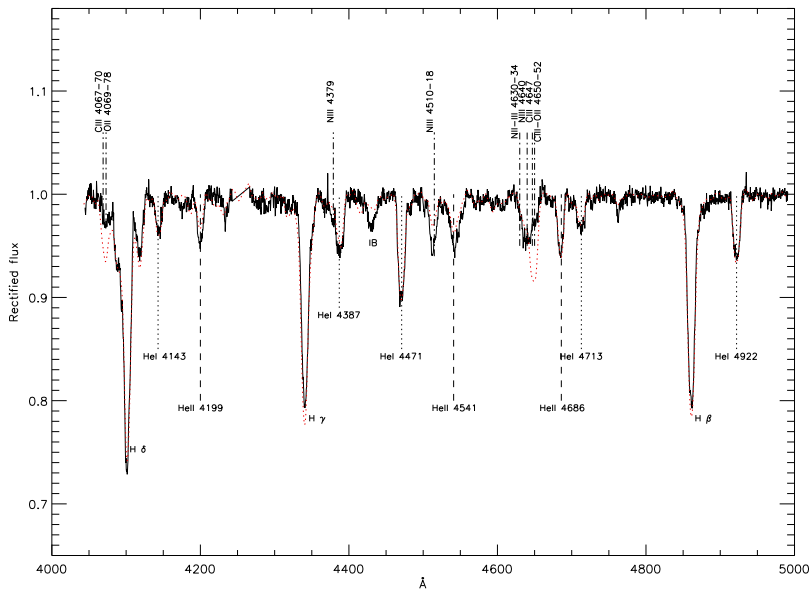}
\caption{Comparison of \object{HD\,149757} ($v{\thinspace}\sin{i}$=400 km\,s$^{-1}$, dotted line) and \object{HD\,191423} ($v{\thinspace}\sin{i}$=450 km\,s$^{-1}$, solid), with the former spectrum degraded to the rotational velocity of the latter. The spectral features used for the CNO analysis are indicated, the CNO blend at $\lambda$\,4630-52 is not used as explained in paper I. IB stands for interstellar band.}\label{rotators}
\end{figure*}

To address the question of the CNO abundances of our target star $\zeta$ Oph we start by comparing its spectrum to that of the fast rotator HD\,191423, analysed in paper I. Rather than comparing them directly we first degrade the spectrum of $\zeta$ Oph, v{\thinspace}sin{\thinspace}$i$=400 km\,s$^{-1}$, up to the higher rotational velocity of HD\,191423 (v{\thinspace}sin{\thinspace}$i$=450 km\,s$^{-1}$, see Fig. \ref{rotators}).

In Fig. \ref{rotators} we see clearly that the \ion{C}{iii}-\ion{O}{ii} features at $\lambda$\,4067-78 of $\zeta$ Oph are stronger than in \object{HD\,191423}, while the \ion{N}{iii} lines at $\lambda$\,4511-24 and $\lambda$\,4379 are weaker. This points to less CNO contamination in our target object.

Since the stellar parameters of both objects are quite similar (see Table 2), we can appreciate how different their CNO content can be. But of course, as they are not exactly the same, the differences in the spectral features will not only be due to a different CNO composition. Therefore this comparison is just illustrative; only after the full CNO analysis can a quantitative comparison be done.

\subsection{Methodology and new He abundance}\label{method}

\begin{figure}
\centering
\includegraphics[width=9cm]{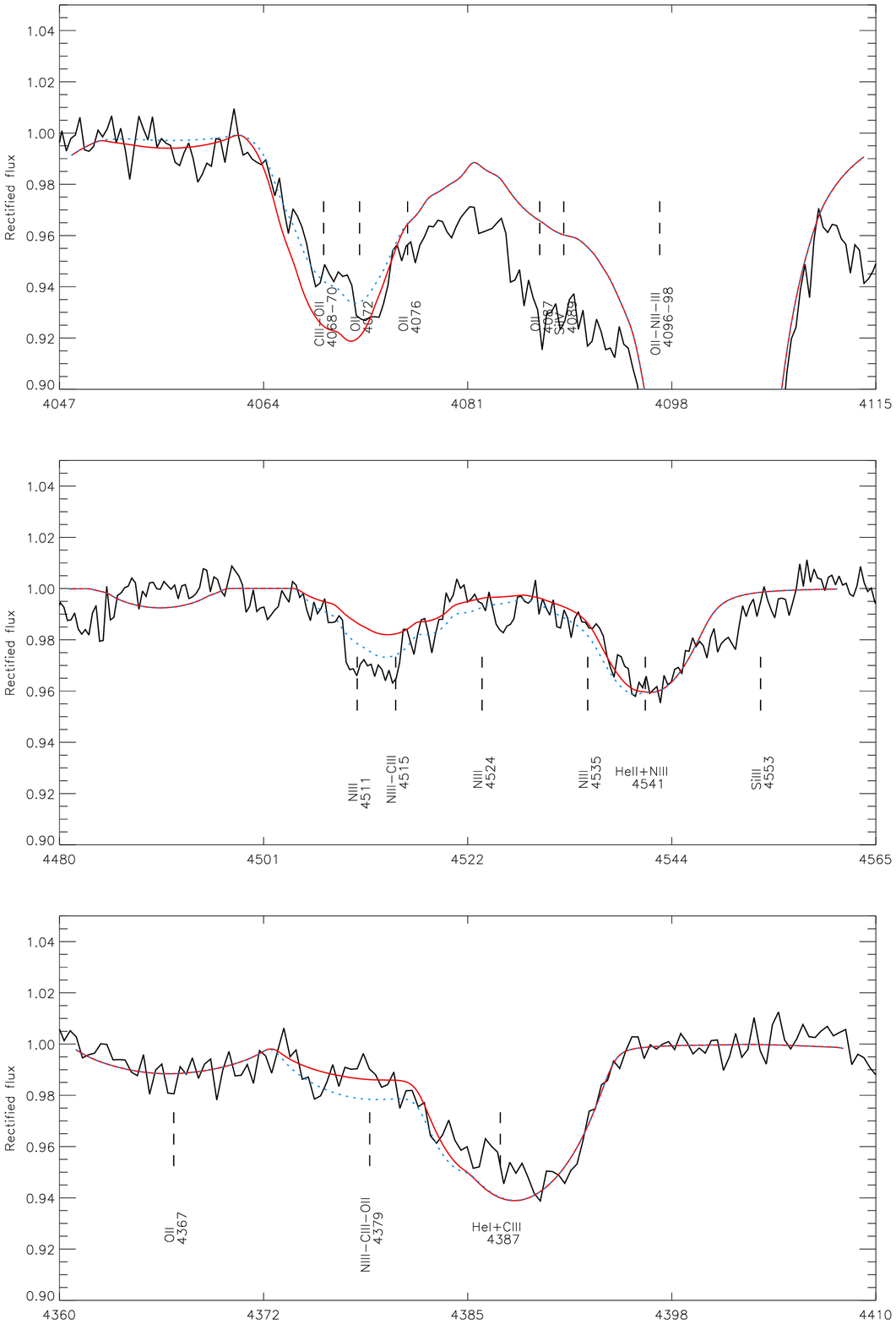}
\caption{H/He \& CNO model spectra to determine the abundances of $\zeta$ Oph. First a model with non-contaminated CNO composition is computed (M6, solid line), and the final best fit to the CNO spectral features yields the stellar abundances (model M7, dotted). Both models are for $\epsilon$(He)=0.11, $\xi$=15
 km\,s$^{-1}$ and the initial CNO model atom (see text).}\label{CNOfits}
\end{figure}

The methodology followed to analyse broad-lined spectra is not the curve of growth method, but direct CNO spectral synthesis. Briefly, we compute a full H/He \& CNO spectrum based on the stellar parameters of our object and with the reference CNO abundances of non-contaminated objects, to compare to the observed spectrum (model M6). We then compute subsequent model spectra until we find the best fit to the observed CNO features (for more details see paper I), that in this case yields $\epsilon$(C)=7.86$\pm$0.3, $\epsilon$(N)=8.34$\pm$0.3 and $\epsilon$(O)=8.69$\pm$0.3 in the standard $\epsilon$(X)=12+$\log \frac{X}{H}$ (model M7, see Fig. \ref{CNOfits}).

The H/He lines are now to be fitted. The previous H/He analysis did not consider any metal species for the synthesis of the spectra, so the contribution of contiguous metallic lines to the H/He lines was neglected.

\begin{figure}
\centering
\includegraphics[width=9cm]{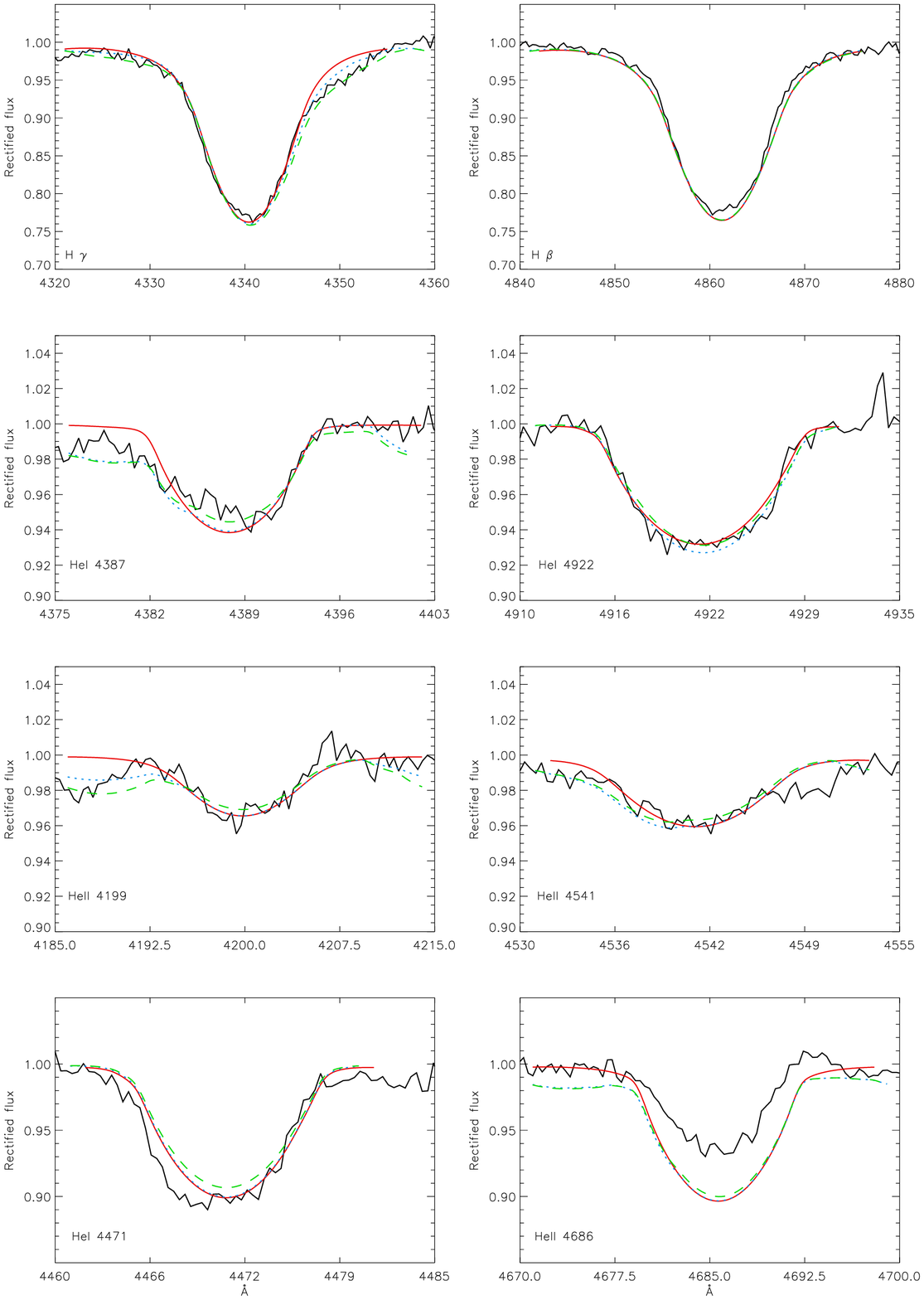}
\caption{Fitting the H/He lines of $\zeta$ Oph when their blends with CNO lines are considered. The H/He model spectrum M5 is shown (solid line) together with the H/He \& CNO models M7 (dotted) and M8 (dashed), with $\epsilon$(He)=0.11 and $\epsilon$(He)=0.09 respectively. Both M7 and M8 are computed for the CNO abundances found for this object: $\epsilon$(C)=7.86$\pm$0.3, $\epsilon$(N)=8.34$\pm$0.3 and $\epsilon$(O)=8.69$\pm$0.3.}
\label{Henew}
\end{figure}

For fast rotators however this contributions can be especially severe, since the lines are so wide. Therefore, as seen in Fig. \ref{Henew}, the synthesis of the full H/He \& CNO spectrum yields a slight correction of the He abundance, from 0.11 to 0.09. The best H/He model M5 (solid line) turns into model M7 (dotted) when the combined H/He \& CNO spectrum is synthesized (for the CNO abundances previously found), and the correction of the He abundance to 0.09 produces better fits to the He lines (model M8, dashed).

The He abundance is determinant in the model atmosphere, that is held fixed for the subsequent line formation of the CNO model spectra. Therefore this change in the He abundance, although small, may cause a change in the model CNO lines. Therefore we compute a new H/He \& CNO model spectrum with the previous CNO abundances on the basis of the new He abundance, and we find that the new fits to the CNO features (model M9, see Fig. \ref{CNOfits2}) are better than before, so the CNO abundances need not be changed. These better fits are due also to the improvements we have implemented in the combined CNO model atom used this time for the spectral synthesis (see sect. \ref{cnoabs}).

\subsection{CNO abundances and improvements in our CNO model atom}\label{cnoabs}
The CNO abundances of our object are obtained as explained in the previous section by fitting the CNO spectral features found in its spectrum. In Fig. \ref{CNOfits} we can see the models computed for $\zeta$ Oph, yielding the following CNO abundances: $\epsilon$(C)=7.86$\pm$0.3, $\epsilon$(N)=8.34$\pm$0.3 and $\epsilon$(O)=8.69$\pm$0.3.

\begin{figure}
\centering
\includegraphics[width=9cm]{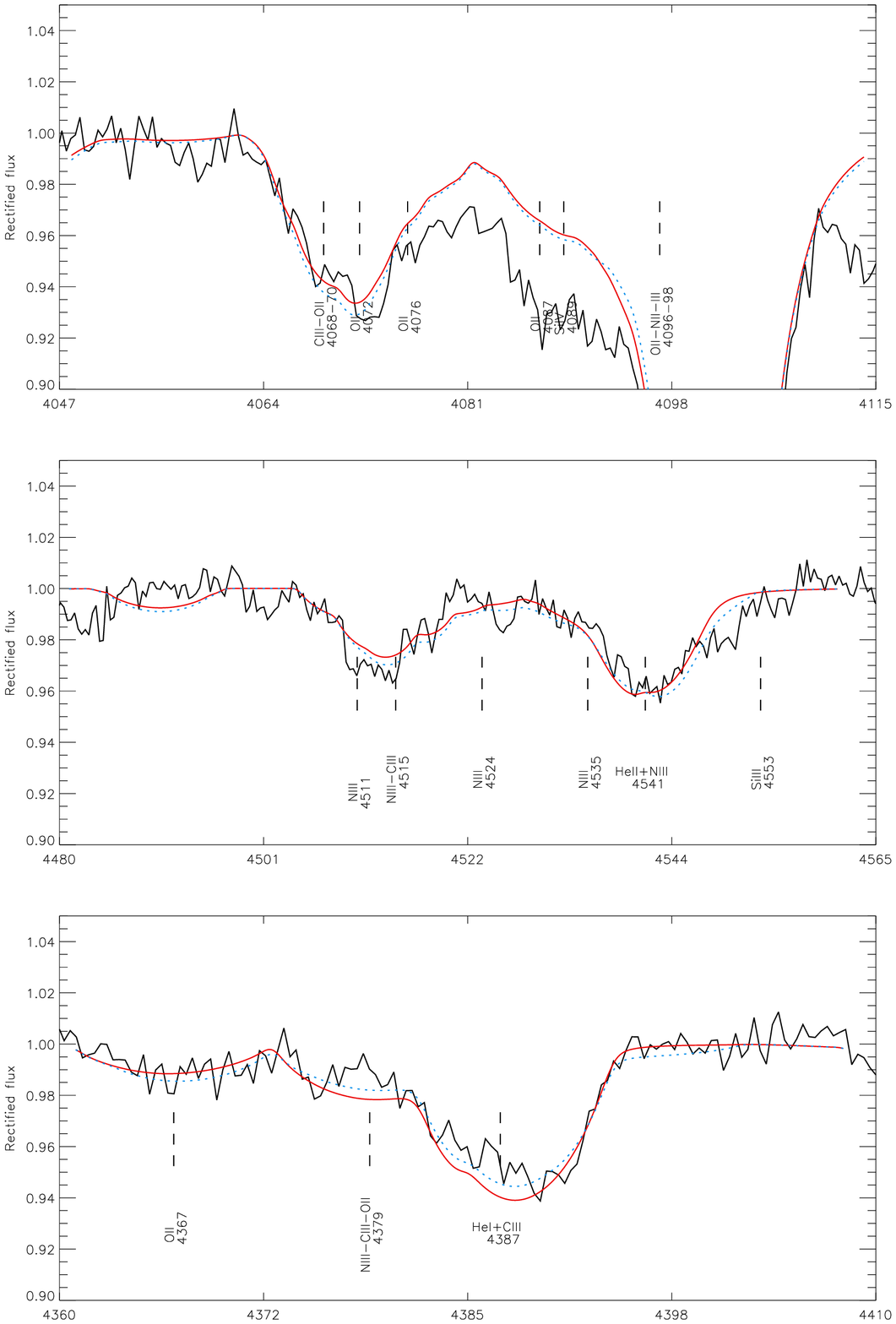}
\caption{Fitting the CNO spectral features of $\zeta$ Oph with models M7 (solid line) and M9 (dotted), both for $\epsilon$(C)=7.86, $\epsilon$(N)=8.34 and $\epsilon$(O)=8.69. M7 is for $\epsilon$(He)=0.11 and the initial CNO model atom and M9 is for $\epsilon$(He)=0.09 and the improved one (see text).}\label{CNOfits2}
\end{figure}

As explained there the determination of the He and CNO abundances is coupled. See in Fig. \ref{CNOfits2} how the small change in the He abundance from 0.11 to 0.09 due to the consideration of the blends with CNO of the H/He lines, does not imply a change in the determined CNO abundances.

In this figure we show not only the effect of the He abundance on the model CNO lines, but also the effect of the new combined CNO model atom used to calculate model M9. In paper I we could only determine a lower limit for the N abundance of the fast rotator HD\,191423, because a fit to all the \ion{N}{iii} features was not possible with a unique value of the N abundance. Therefore we have improved our combined CNO model atom in the spectral range around \ion{He}{ii}\,4541 and \ion{N}{iii}\,4511-24, and around \ion{He}{i}\,4387 and \ion{N}{iii}\,4379 in order to overcome this difficulty.

\begin{table}[!hb]
\label{newtrans}
\caption[ ]{New transitions included in the CNO model atom. The level energies are taken from the NIST database (http://physics.nist.gov), in cm$^{-1}$, and the $\log$\,gf values from VALD. \ion{N}{iii}\,4379 and \ion{C}{iii}\,4056 were already in the model atom but we have updated their $\log$\,gf to that of VALD}
\begin{tabular}{lccc}
\hline
Trans. & E$_l$ & E$_u$ & $\log$\,gf \\
\hline
\ion{C}{iii}\,4056 & 324212.49 & 348859.99 & 0.199 \\
\hline
\ion{N}{iii}\,4379 & 320365.7  & 343116.4  &  1.010 \\
\ion{N}{iii}\,4527 & 287591.5  & 309691.8  & -0.471 \\
\ion{N}{iii}\,4530 & 287591.5  & 309656.2  & -1.161 \\
\ion{N}{iii}\,4539 & 311690.3  & 333712.0  & -0.453 \\
\ion{N}{iii}\,4544 & 311715.2  & 333712.0  & -0.152 \\
\ion{N}{iii}\,4546 & 314217.3  & 336206.9  &  0.004 \\
\ion{N}{iii}\,4547 & 287706.9  & 309691.8  & -1.384 \\
\hline
\end{tabular}
\end{table}

The improvements in each region consisted of including all the \ion{C}{iii}, \ion{N}{iii} and \ion{O}{ii} transitions of the VALD database not already present in the model atom and a revision of the oscillator strengths for consistency with those of VALD (Kupka et al. \cite{Kupka99}, Ryabchikova et al. \cite{Ryab99}). 

In Table 3 we give all the details. The determination of the N abundance seems to be resolved, at least for this case.


\section{Comparison with evolutionary models}\label{evol}

In the HR diagram in Fig. \ref{149HRD} we see how the evolutionary tracks have evolved from Meynet \& Maeder (\cite{meymae00}) to Meynet \& Maeder (\cite{meymae03}). The basic differences between the two sets of models are the use of the mass loss rates of Vink et al. (\cite{vink00}, \cite{vink01}) and Nugis \& Lamers (\cite{nugis00}) for O and WR stars and the consideration of the wind anisotropies induced by rotation, both included in the 2003 models. Some other improvements are also introduced, see Meynet \& Maeder (\cite{meymae03}) for details.

\begin{figure}
\centering
\includegraphics[width=9cm]{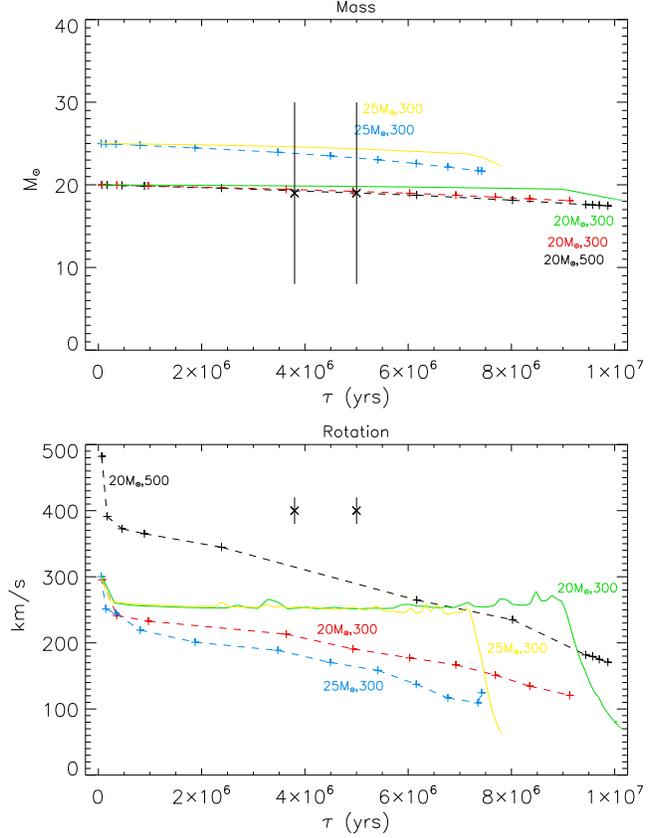}
\caption{Comparison of our measured mass and (projected) rotational velocity with the MS evolutionary models of Meynet \& Maeder (\cite{meymae00}, dashed line) and Meynet \& Maeder (\cite{meymae03}, solid). Models are for the quoted pairs of initial mass and equatorial rotational velocity. We plot the two possible upper limits for the age of our object: 5 and 3.8 Myrs according to the 2000 and 2003 tracks respectively.}\label{massvrot}
\end{figure}

\begin{figure}
\centering
\includegraphics[width=9cm]{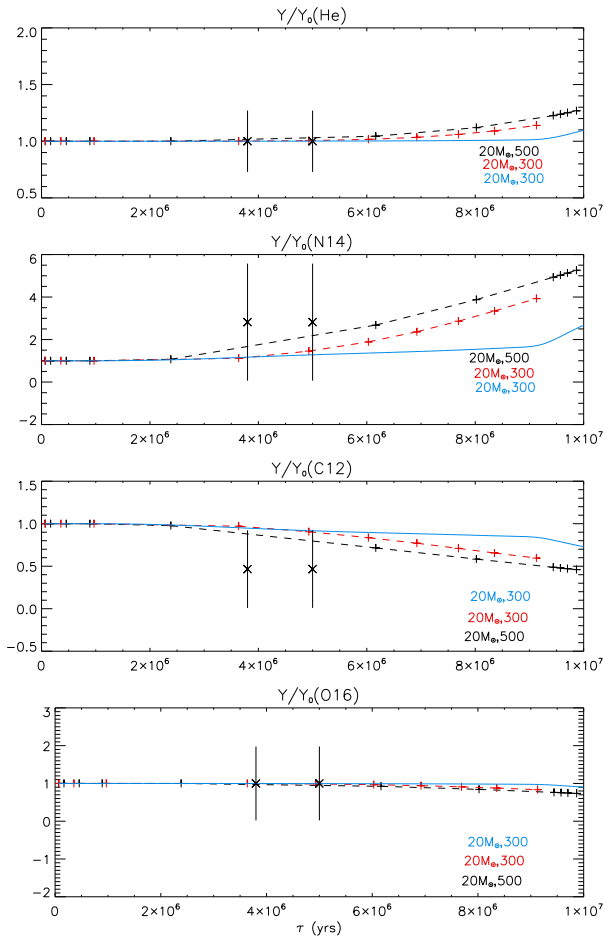}
\caption{As in Fig.\ref{massvrot} but for the He \& CNO abundances. Note that our reference mass fractions, Y(X)$_0$, are those of the reference non-contaminated objects while the initial composition in the evolutionary models is solar.}\label{hecno}
\end{figure}

The consequences of these differences in the MS are wider tracks at slightly lower luminosities and longer MS lifetimes. Concerning the evolution of the rotational velocities and surface composition (see Figs. \ref{massvrot} and \ref{hecno}) the 2003 models show smoother changes with stellar age: rotation decreases considerably more slowly and the He \& CNO enrichments/depletions appear later and are fewer.

We compare first our mass and (projected) rotational velocity, see Fig. \ref{massvrot}. Our mass of 19 M$_{\odot}$ is in agreement with both the 2000 and 2003 values for the models with initial 20 M$_{\odot}$ mass, although the star's position in the HRD lies closer to the 25 M$_{\odot}$ tracks. This effect is related to the well-know ``mass discrepancy'' (Herrero et al. \cite{h92}) that is still not completely solved (Repolust et al. \cite{repo04}). Our value of 400 km\,s$^{-1}$ for the projected rotational velocity is much too high for the 2000 models while a 2003 model starting at $\approx$ 500 km\,s$^{-1}$ would be able to reproduce our measurement. 

Concerning the surface composition (see Fig. \ref{hecno}), our He and O abundances are in agreement with the evolutionary predictions, that hardly change in the MS, while our N enrichment and C depletion are much higher than predicted. The comparison is worse with the 2003 models.

Thus, under the 2003 tracks, our high rotational velocity could be explained, but our N enrichment and C depletion are much too high. The 2000 tracks on the contrary cannot explain our high projected rotational velocity, but predict our observed abundances of He, N and O.

This may indicate that either the mixing mechanisms are more efficient than considered in the evolutionary models for a given rotational velocity, or that the loss of angular momentum is less efficient so that high rotational velocities last longer in the MS. This is what Herrero \& Lennon (\cite{hl04}) find for O stars, and Trundle et al. (\cite{tru04}) for B supergiants.

However, until the runaway nature of $\zeta$ Oph is clarified (in a forthcoming paper) the results presented here cannot serve as constraints for the evolutionary models, since the origin of our object may be a close binary system rather than a single massive star.


\section{Conclusions}\label{discuss}

We find that the fast rotator $\zeta$ Oph shows a surface composition with N enriched and C depleted, with normal He and O abundances. Our abundances are based on our last analysis of this object, that includes improvements in our method first presented in paper I. In particular we are now capable of giving more reliable N abundances due to the improvements introduced in the model atom used for the analysis.

When comparing our surface composition with the predictions of the evolutionary models of the Geneva group (Meynet \& Maeder, \cite{meymae00}, \cite{meymae03}) we find better (although not complete) agreement with the 2000 models, while for the rotational velocity only the 2003 models would produce the high value of 400 km\,s$^{-1}$ we observe.

The fact that $\zeta$ Oph is a runaway prevents us from going any further in discussing the meaning of these disagreements, that may be just indicating that this object does not come from the evolution of a single massive star, but from the evolution of a close binary system. Only after the runaway nature of $\zeta$ Oph is clarified can this discussion continue.


\begin{acknowledgements}
We want to thank D.J. Lennon and T. Augusteijn for many valuable discussions and encouragement at the beginning of this work. G. Meynet has also contributed significantly with his clear explanations to our understanding of their evolutionary models. This work has been partly supported by the Spanish MEC under project PNAYA 2001-0436. We want to thank also I.D. Howarth and the A\&A language editor for their careful revision of the text.
\end{acknowledgements}

\end{document}